\documentclass[11pt]{emulatenat}

\newcommand{\isotope}[2]{\ensuremath{\mathrm{^{#1}#2}}}
\newcommand{\carbon}[1][12]{\isotope{#1}{C}}
\newcommand{\sodium}[1][23]{\isotope{#1}{Na}}
\newcommand{\magnesium}[1][24]{\isotope{#1}{Mg}}
\newcommand{\aluminum}[1][27]{\isotope{#1}{Al}}
\newcommand{\calcium}[1][40]{\isotope{#1}{Ca}}
\newcommand{\scandium}[1][45]{\isotope{#1}{Sc}}
\newcommand{\titanium}[1][48]{\isotope{#1}{Ti}}
\newcommand{\vanadium}[1][51]{\isotope{#1}{V}}
\newcommand{\chromium}[1][52]{\isotope{#1}{Cr}}
\newcommand{\manganese}[1][55]{\isotope{#1}{Mn}}
\newcommand{\iron}[1][56]{\isotope{#1}{Fe}}

\newcommand{\yttrium}[1][89]{\isotope{#1}{Y}}
\newcommand{\rubidium}[1][85]{\isotope{#1}{Rb}}
\newcommand{\strontium}[1][88]{\isotope{#1}{Sr}}
\newcommand{\bromine}[1][79]{\isotope{#1}{Br}}
\newcommand{\krypton}[1][84]{\isotope{#1}{Kr}}
\newcommand{\zirconium}[1][94]{\isotope{#1}{Zr}}

\newcommand{\Qec}{\ensuremath{Q_{\mathrm{EC}}} }
\newcommand{\apj}{Astrophys.~J.}
\newcommand{\apss}{Astrophys.~J. Suppl.}
\newcommand{\aplett}{Astrophys.~J. Lett.}
\newcommand{\apjl}{\aplett}
\newcommand{\araa}{Ann.~Rev.~Astron.~\& Astrophys.} 
\newcommand{\AsAs}{Astron.~\& Astrophys.} 
\newcommand{\prc}{Phys.~Rev.~C}
\newcommand{\npa}{Nucl. Phys. A}
\newcommand\physrep{Phys.~Rep.}

\newcommand{\ECb}{electron-capture/$\beta^-$-decay} 
\newcommand*{\dif}{\ensuremath{\mathrm{d}}}
\newcommand*{\enu}{\ensuremath{\varepsilon_{\nu}}}

\renewcommand*{\figurename}{Figure}
\renewcommand*{\tablename}{Table}

\title{Strong neutrino cooling by cycles of electron capture and $\beta^-$ decay in
neutron star crusts}
 \author{%
 H. Schatz$^{1,2,3}$, 
 S. Gupta$^{4}$,
 P. M{\"o}ller$^{2,5}$,
 M. Beard$^{2,6}$,
 E.F. Brown$^{1,2,3}$,
 A.T. Deibel$^{2,3}$,
 L.R. Gasques$^{7}$,\\
 W.R. Hix$^{8,9}$, 
 L. Keek$^{1,2,3}$,
 R. Lau$^{1,2,3}$,
 A.W. Steiner$^{2,10}$,
 M. Wiescher$^{2,6}$
 }
\date{}
 
\begin{document}
\maketitle  
\begin{affiliations}
  \item National Superconducting Cyclotron Laboratory, Michigan State University, 640 S. Shaw Lane, East Lansing, Michigan 48824, USA
  \item Joint Institute for Nuclear Astrophysics, 225 Nieuwland Science Hall, University of Notre Dame, Notre Dame, Indiana 46556, USA
  \item Department of Physics and Astronomy, Michigan State University, 567 Wilson Road, East Lansing, Michigan 48824, USA
  \item Indian Institute of Technology Ropar, Nangal Road, Rupnagar (Ropar), Punjab 140 001, India
  \item Theoretical Division, MS B214, Los Alamos National Laboratory, Los Alamos, New Mexico 87545, USA
  \item Department of Physics, 225 Nieuwland Science Hall, University of Notre Dame, Notre Dame, Indiana 46556, USA
  \item Departamento de F\'{i}sica Nuclear, Instituto de F\'{i}sica da Universidade de S\~{a}o Paulo, Caixa Postal 66318, 05315-970 S\~{a}o Paulo, SP, Brazil
  \item Physics Division, Oak Ridge National Laboratory, P.O. Box 2008, Oak Ridge, TN 37831-6354, USA
  \item Department of Physics and Astronomy, University of Tennessee, 401 Nielsen Physics Building,
  1408 Circle Drive, Knoxville, TN 37996-1200,  USA
  \item Institute for Nuclear Theory, University of Washington, Physics/Astronomy Building,
  Box 351550, Seattle, WA 98195-1550, USA
\end{affiliations}
  
\begin{abstract}
The temperature in the crust of an accreting neutron star, which comprises its outermost kilometer,  is set by heating from nuclear reactions at large densities\cite{sato79,Haensel1990,Gupta2007,Haensel2008}, neutrino cooling\cite{Yakovlev2001,Steiner2009}, and heat transport from the interior\cite{Cumming2001,Strohmayer2002,Brown2004,Cumming2006,Keek2011}. The heated crust has been thought to affect observable phenomena at shallower depths, such as thermonuclear bursts in the accreted envelope\cite{Cumming2006,Keek2011}. Here we report that cycles of electron capture and its inverse, $\beta^-$ decay, involving neutron-rich nuclei at a typical depth of about 150~m, cool the outer neutron star crust by emitting neutrinos while also thermally decoupling the surface layers from the deeper crust. This ``Urca'' mechanism\cite{Gamow1941-Urca} has been studied in the context of white dwarfs\cite{Tsuruta1970} and Type Ia supernovae\cite{Paczynski1972, Woosley1986}, but hitherto was not considered in neutron stars, because previous models\cite{sato79,Haensel1990} computed the crust reactions using a zero-temperature approximation and assumed that only a single nuclear species was present at any given depth. This thermal decoupling means that X-ray bursts and other surface phenomena are largely independent of the strength of deep crustal heating. The unexpectedly short recurrence times, of the order of years, observed for very energetic thermonuclear superbursts\cite{Keek2008} are therefore not an indicator of a hot crust, but may point instead to an unknown local heating mechanism near the neutron star surface.
\end{abstract}

\begin{figure}
\centerline{\includegraphics[width=\figwidth]{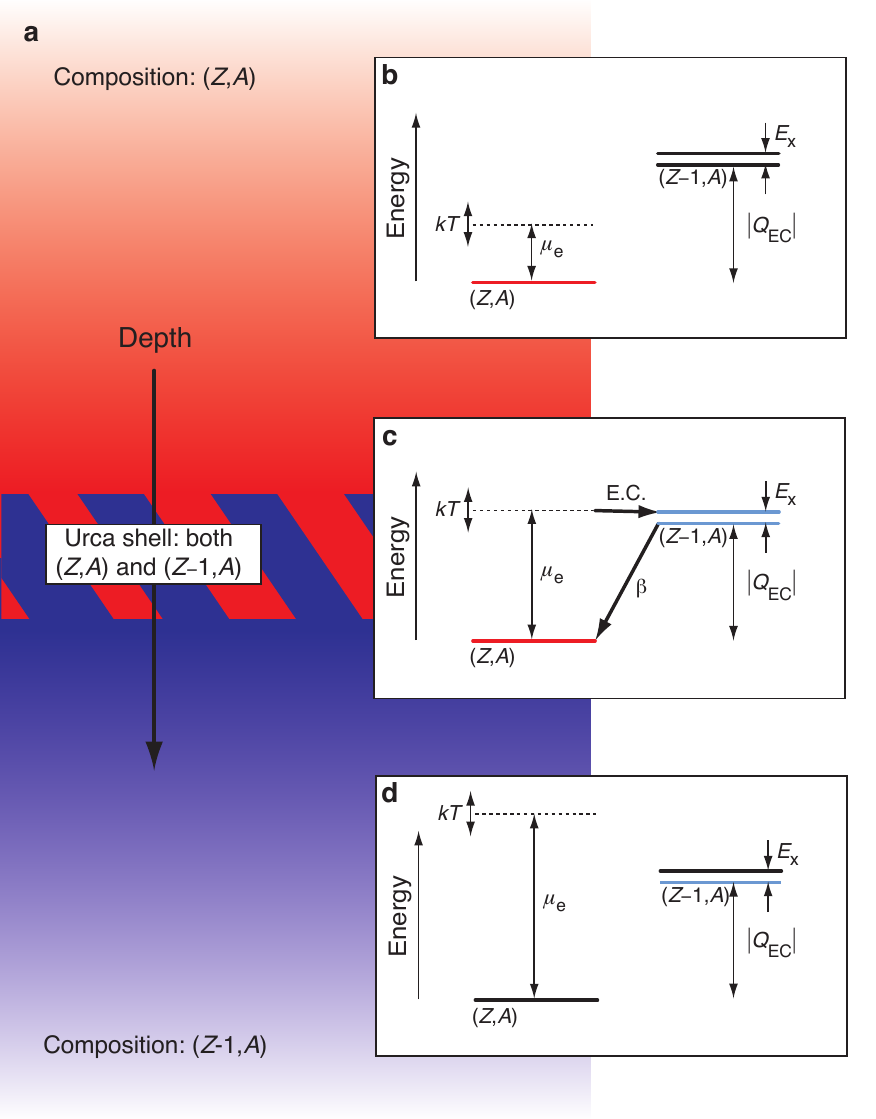}}
\caption{\label{Fig_Schematic} 
\textbf{Schematic nuclear energy-level diagrams for an electron-capture/$\boldsymbol{\beta^{-}}$-decay pair.}
\textbf{a}, Illustration of compositional layers in the neutron star crust; \textbf{b--d}, energy level diagrams. 
In the shallow region above the Urca shell, where the nuclear composition has charge number $Z$ and mass number $A$, $(Z,A)$, the electron chemical potential $\mu_{e}$ is less than $|\Qec|$, the energy threshold for electron capture, and electron capture is energetically blocked (\textbf{b}). In the deeper region below the Urca shell, $\mu_{e} > |\Qec|$: electron capture has therefore occurred, the composition consists of nuclei $(Z-1,A)$, and the degenerate electrons block the phase space for electron emission via $\beta^-$ decay (\textbf{d}). In the Urca shell between these regions, $\mu_{e}\approx |\Qec|$.  As a result, both electron capture and $\beta^-$ decay are possible  (\textbf{c}), and rapid cycling between the nuclei $(Z,A)$ and $(Z-1,A)$ leads to a strong neutrino emissivity.
}
\end{figure}

Continual accretion onto a neutron star pushes the ashes of surface thermonuclear burning, which is often observed as Type I X-ray bursts\cite{Woosley2004a,Galloway2008}, to greater pressures and densities where the nuclei form a rigid lattice\cite{Horowitz2009} known as the crust. With increasing depth, these ashes are transformed by capture of degenerate electrons into increasingly neutron-rich nuclei\cite{sato79,Haensel1990,Gupta2007,Haensel2008}. 
A particular electron-capture reaction $(Z,A) + e^- \rightarrow (Z-1,A) +  {\nu}_e$ from a parent nucleus $(Z,A)$ with charge number $Z$ and mass number $A$ to a daughter nucleus $(Z-1,A)$ occurs  at a well-defined depth, where the electron chemical potential $\mu_e \approx |\Qec|+E_x$. Here $\Qec$ is the 
(negative) electron-capture Q-value (the difference between the parent and daughter ground-state masses and hence the energy needed for the reaction to occur) and $E_{x}$ is the excitation energy of the lowest state in the daughter nucleus that can be populated by electron capture.  In the commonly used zero-temperature approximation, the reverse $\beta^{-}$-decay reaction $(Z-1,A) \rightarrow (Z,A) + e^- + \bar{\nu}_e$ is blocked because there is no phase space available in which to re-emit the captured electron. 
At finite temperature and for $E_x <  kT$, however, $\beta^-$ decay via the re-emission of 
an electron with an energy close to $|\Qec|$ is not completely blocked. As a result, the boundary between a layer containing nuclei $(Z,A)$ and a deeper layer containing $(Z-1,A)$ is  a shell with mixed composition spanning a range of electron chemical potential
$|\Qec| - kT \lesssim \mu_e \lesssim |\Qec| + kT$ that corresponds to a thickness of a few meters within the neutron star crust. Inside this shell both electron capture and its inverse $\beta^-$ decay occur (see Fig.~\ref{Fig_Schematic}). If these reactions cycle back-and-forth rapidly, the result is a strong neutrino emission, known as an \emph{Urca} process\cite{Gamow1941-Urca}, that cools the neutron star crust. Such Urca shells have been studied before, both in the context of white dwarfs\cite{Tsuruta1970}, Type Ia supernovae\cite{Paczynski1972, Woosley1986}, and stellar ONeMg cores producing electron-capture supernovae\cite{Jones2013}. The effect has not been considered in the context of accreting neutron stars. Most earlier models of accreted crusts\cite{sato79,Haensel1990,Haensel2008} were computed in the zero-temperature limit, in which the composition switches sharply at the energetic thresholds with no available phase space for cycling; indeed, in this limit, only one nuclide of a reaction pair is present at a given depth. Urca shell cooling relies on phase space unblocking at finite temperature and the presence of both reaction pair nuclides in the shell. More recent reaction network calculations\cite{Gupta2007} did not include $\beta^-$-decays as they were not considered to be important, and any Urca cycling was estimated to be negligible.  The importance of Urca shell cooling is revealed here through the use of a full reaction network that includes both electron capture and $\beta^-$ decay on an equal footing, that takes into account the rates of subsequent reactions that deplete the \ECb\ pairs, and that follows the evolution in time of a fluid element as it is pushed through a reaction shell.

\begin{figure}
\centerline{\includegraphics[width=\figwidth]{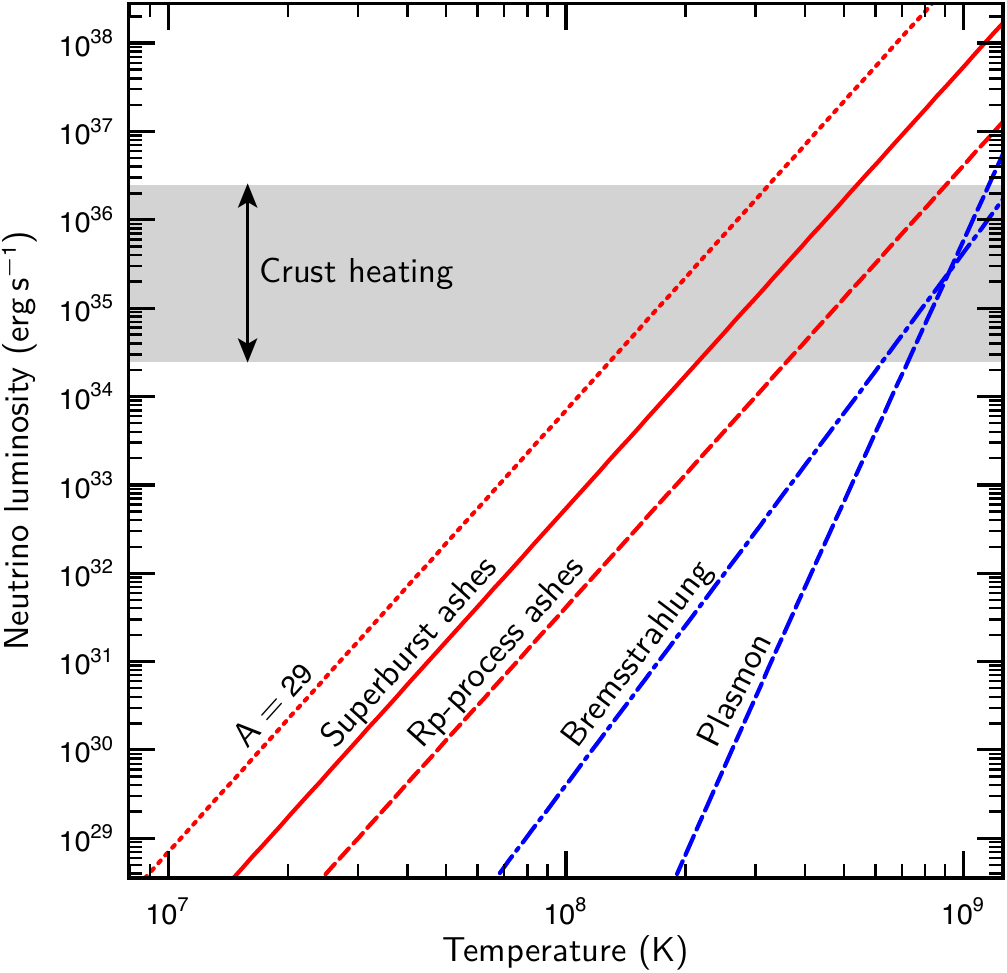}}
\caption{\label{Fig_Rates}
\textbf{Neutrino luminosities in the accreted neutron star crust.}  For purposes of comparison, estimates for the neutrino luminosities were obtained by integrating the corresponding emissivities over a neutron star crust with a constant local temperature, following the methodology of Ref.\ \citen{Cumming2006}.  Although an actual neutron star crust is not isothermal, the temperature variation across the crust is not large (typically less than a factor of 2). 
The Urca shell luminosities were calculated for superburst ashes\cite{Keek2012} (solid red line), X-ray burst ashes produced by the rapid proton capture process\cite{Woosley2004a}  (dashed red line), and a pure $A=29$ composition (dotted red line) to demonstrate the maximum effect. 
Also shown are the neutrino luminosities from electron-nucleus bremsstrahlung (blue dot-dashed line) and plasmon decay (blue dashed line).
For 
comparison, we show estimates for the total crustal heating (shaded band).  This is based on a local heating rate of $(1.9\,\mathrm{MeV/u})\times\dot{M}$ for a range of accretion rates $10^{16}\,\mathrm{g/s}\lesssim \dot{M} \lesssim 10^{18}\,\mathrm{g/s}$, which is representative for observed neutron stars. Urca shells dominate the neutrino luminosity from the crust and can balance the crust heating at moderate temperatures $\gtrsim 2\times 10^{8}\,\mathrm{K}$. The temperature scaling assumes $E_x \ll kT$ (see the Supplementary Information section 1). 
}
\end{figure}

In order for this cycling of electron capture and $\beta^{-}$ decays between two nuclear species to occur, the nuclei involved must satisfy two conditions. First, the transitions must proceed between low-lying states ($E_x \lesssim kT$ is required for both the electron capture and the $\beta^-$ decay). 
In addition, within an \ECb\ pair, the $\beta^{-}$ decaying nucleus must not have a strong electron-capture branch, as these electron captures would remove nuclei from the Urca-cycle thereby reducing its effect or eliminating it entirely. 
The cooling rate depends on the strength of the transition (the $ft$-value, which is directly related to the matrix element connecting the parent and daughter states) and the energy threshold; the integration over the available phase space produces a characteristic $T^5$ scaling with temperature\cite{Tsuruta1970}. 
The formation of Urca shells with large strength is enabled by strong nuclear deformations that tend to spread nuclear electron-capture strength to lower excited states, thereby lowering $E_x$\cite{Krumlinde1984} (see Extended Data Fig.~\ref{Fig_Nilsson}). There are a number of \ECb\ pairs that fulfill these conditions for forming fast-cooling Urca shells (see Tab.~\ref{Tab_top}). 
The degree to which these shells are activated in a neutron star crust depends on the initial composition produced by thermonuclear burning on the neutron star surface. Because electron captures in the outer crust preserve the mass number $A$, the abundance of an \ECb\ pair, and therefore its absolute neutrino luminosity, is set by the abundance of nuclei with the same mass number in the ashes of the surface thermonuclear burning. As is evident from Fig.~\ref{Fig_Rates}, neutrino cooling by Urca shells is by far the dominant neutrino emission process in the crust for typical crust compositions.


\begin{figure}
\centerline{\includegraphics[width=\doublewide]{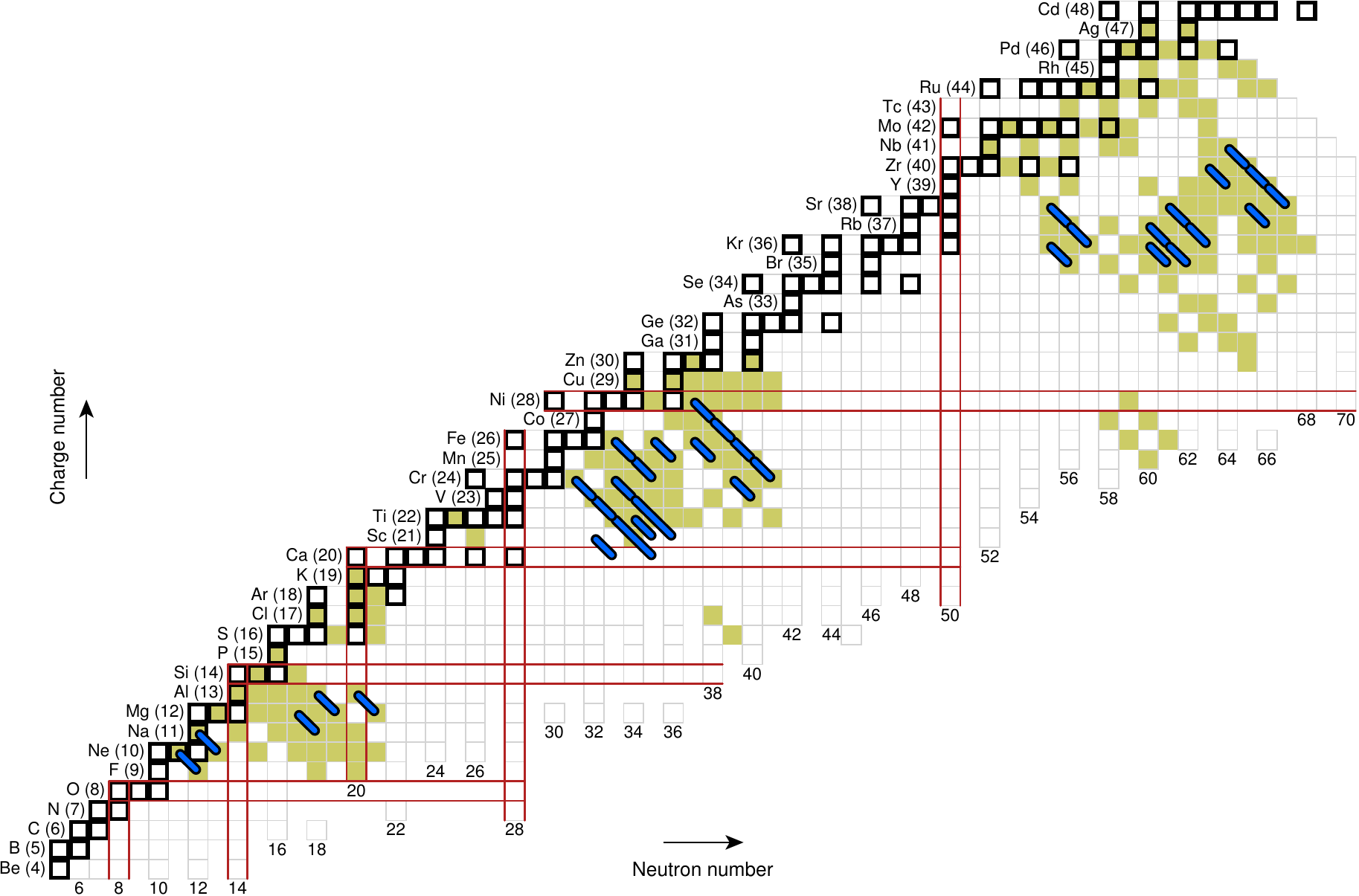}}
\caption{\label{Fig_Chart} 
\textbf{Electron-capture/$\boldsymbol{\beta^{-}}$-decay pairs on a chart of the nuclides.} 
The thick blue lines denote \ECb\ pairs that would generate a strong neutrino luminosity in excess of $5\times10^{34}\,\mathrm{erg\,s^{-1}}$ at $T=0.51\,\mathrm{GK}$ for a composition consisting entirely of the 
respective \ECb\ pair.
The locations of the strong \ECb\ cooling pairs on the chart of nuclides largely coincide with regions where allowed electron-capture and 
$\beta^-$-decay transitions are predicted to populate low-lying states and subsequent electron capture is blocked (shaded squares, see also the discussion in Ref.\ \citen{Gupta2007}). These are mostly regions between the closed neutron and proton shells (pairs of horizontal and vertical red lines),
where nuclei are significantly deformed (see the Supplementary Information section 4).  Nuclides that are $\beta^{-}$-stable under terrestrial conditions are shown as squares bordered by thicker lines.
Nuclear charge numbers are indicated next to element symbols. 
} 
\end{figure}



The greatly enhanced crust neutrino emissivity at rather shallow depths changes the long-standing assumption that rapidly accreting neutron stars have a significant luminosity from deep crustal heating, which directly influences thermonuclear burning in their accreted envelopes. 
In the absence of crust Urca shells,
if the crust has a low thermal conductivity, and if the core neutrino emissivity is weak, then deep crustal heating would generate significant heat flow towards the neutron star surface\cite{brown:nuclear}. Models of thermonuclear bursts\cite{Woosley2004a,Keek2011} use this emergent luminosity from the deep crust as a boundary condition, which sets in part the ignition density and temperature. The presence of strongly temperature-sensitive Urca cycles limits the temperature at the location of the shells, however, and may even require an inward directed luminosity from the accreted envelope.  
Even for conditions in the deep crust that are favorable for sending a large heat flux to the surface, the Urca shells reemit this heat as neutrinos preventing it from reaching the surface layers (see Supplementary Information section 3).

To establish the robustness of our conclusions 
with respect to nuclear physics uncertainties, we used two different mass models, namely the
FRDM\cite{Moller1995} and the HFB-21\cite{Goriely2010} mass model.  The use of FRDM masses instead of those from HFB-21 reduces the Urca shell neutrino luminosity by 90\% for a superburst ash composition (see the Supplementary Information S5 for details). For both mass models the temperature at the superburst ignition depth is  $< 5\times 10^{8}\,\mathrm{K}$ if the Urca shell neutrino emissivity is included; and for both mass models the temperature has a significant local minimum at the location of the Urca shell (see Extended Data Fig.~\ref{f.crust_temps}). 

This has strong implications for the ignition of superbursts, which are thought to be triggered by the unstable thermonuclear reaction $\carbon+\carbon$. 
The observed recurrence times, of the order of one year, are much shorter than predictions of current models\cite{Cumming2001,Strohmayer2002,Cumming2006}, which  indicates that the temperature at the ignition depth is underestimated. 
The presence of Urca shells implies that this observation does not point to an unexpectedly hot crust\cite{Gupta2007,Cumming2006}.  Because of the Urca shells, we conclude that the standard carbon-ignition scenario for superbursts requires a powerful, as yet unknown, heat source that operates at surprisingly shallow densities $\lesssim 10^{10}\,\mathrm{g\,cm^{-3}}$ very near the carbon ignition layer. Alternatively other, more exotic, mechanisms would have to be found to ignite and power superbursts. 

Urca shell cooling therefore forces fundamental changes in current superburst models. Realistic ignition conditions must now include a strong localized heat source at a depth close to that of ignition, with a strong heat flux flowing inwards into the Urca shell cooling layer.  The resulting temperature profile at ignition is therefore dramatically different from that assumed previously, which may alter predicted light curves. In addition, during the explosion, the temperature at the ignition depth rises to $\approx 10^{9}\,\mathrm{K}$.  Heat from this layer will diffuse inward; based on the thermal diffusion timescale in the neutron star crust\cite{Brown2009} we estimate that in the absence of any neutrino emission, the temperature would rise to $\approx 10^{9}\,\mathrm{K}$ at the depth of the Urca shell within a day following ignition.  The presence of a strong ``heat sink'' at that depth, however, will prevent the deeper layers from rising in temperature and will, therefore, force the observed light curve to decay faster than expected over timescales of roughly  one day.  
Current superburst observations on this timescale are rare and provide data of limited quality\cite{Keek2012}, but a dedicated program
of superburst follow-up observations with current instrumentation could address this problem. Detailed simulations, which are beyond the scope of this paper, are required to quantify the effect of Urca shell cooling on superburst light curves. Such simulations must be based on revised 
superburst models that  can incorporate the new ignition conditions consistent with Urca shell cooling.

Another observational signature might be found in neutron star cooling following an accretion outburst.
Unlike crustal heating, the rate of Urca shell cooling does not scale with accretion rate, but rather depends only on temperature.  Cooling will therefore continue in transiently accreting neutron stars once accretion has turned off,
and might affect observations of the cooling crusts in the hottest of these systems, such as XTE~J1701$-$462,\cite{Page2013Forecasting-neu} as the Urca shell neutrino cooling rate is comparable to typical photon 
luminosities of $10^{32}\textrm{--}10^{33}\,\mathrm{erg/s}$ for typical initial crust temperatures of $1\textrm{--}3\times 10^8\,\mathrm{K}$.

\begin{addendum}
 \item This project was funded by NSF grants PHY 08-22648 (Joint Institute for Nuclear Astrophysics) and PHY 06-06007. A.S. is supported by INT DOE grant DE-FG02-00ER41132. EFB is supported by NSF grant AST 11-09176. PM is supported by the National Nuclear Security Administration of the U. S. Department of Energy at Los Alamos National Laboratory under Contract No.\ DE-AC52-06NA25396. We thank
 D.M. Yakovlev, P. Shternin, and S. Reddy for discussions and comments on the manuscript. 
 \item [Supplementary Information] is linked to the online version of the paper at
www.nature.com/nature.
\item [Author contributions] H.S. calculated crust models, analyzed data, and prepared the manuscript. S.G. developed and implemented the phase space calculation. S.G. and P.M. calculated the weak transition rates. E.F.B. calculated crust temperature profiles and assisted with writing the manuscript. A.T.D. computed the temperature scaling of the neutrino emission. L.K. calculated superburst models. M.B., W.R.H., and R.L. wrote model code. All authors contributed to the interpretation of the results, and contributed to or commented on the manuscript. 
\item [Reprints and Permissions]  information is available at www.nature.com/reprints
 \item[Competing Interests] The authors declare that they have no
competing financial interests.
 \item[Correspondence] Correspondence and requests for materials
should be addressed to H. Schatz~(email: schatz@nscl.msu.edu).
\end{addendum}

\begin{table}
\sffamily\footnotesize
\renewcommand{\isotope}[2]{\ensuremath{\mathsf{^{#1}#2}}}
\caption{\label{Tab_top} \textbf{Electron-capture/$\boldsymbol{\beta^{-}}$-decay pairs with the largest cooling rates.}}
\begin{center}
\begin{tabular}{llrrr}
\hline\hline
\multicolumn{2}{c}{Electron-capture/$\beta^-$-decay pair} & Density\textsuperscript{\ddag} & Chemical Pot.\textsuperscript{\ddag} & Luminosity\textsuperscript{\dag}\\ 
  Parent       &  Daughter\textsuperscript{\P} & $\mathrm{10^{10}\,g\,cm^{-3}}$ & $\mathrm{MeV}$  & $\mathrm{10^{36}\,erg/s}$\\ 
\hline
\magnesium[29]  & \sodium[29]                   &  4.79 & 13.3 & 24 \\
\titanium[55]   & \scandium[55], \calcium[55]   &  3.73 & 12.1 & 11 \\
\aluminum[31]   & \magnesium[31]                &  3.39 & 11.8 &  8.8 \\
\aluminum[33]   & \magnesium[33]                &  5.19 & 13.4 &  8.3 \\
\titanium[56]   & \scandium[56]                 &  5.57 & 13.8 &  3.5 \\
\chromium[57]    & \vanadium[57]                 &  1.22 &  8.3 &  1.6 \\
\vanadium[57]   & \titanium[57], \scandium[57]  &  2.56 & 10.7 &  1.6 \\
\chromium[63]   & \vanadium[63]                 &  6.82 & 14.7 &  0.97 \\
\zirconium[105] & \yttrium[105]                 &  3.12 & 11.2 &  0.92 \\
\manganese[59]  & \chromium[59]                 &  9.45 &  7.6 &  0.88 \\
\strontium[103] & \rubidium[103]                &  5.30 & 13.3 &  0.65 \\
\krypton[96]    & \bromine[96]                  &  6.40 & 14.3 &  0.65 \\
\iron[65]       & \manganese[65]                &  2.34 & 10.3 &  0.60 \\
\manganese[65]  & \chromium[65]                 &  3.55 & 11.7 &  0.46 \\
\hline
\end{tabular}\end{center}
\scriptsize
\P\ The listing of two electron-capture daughter isotopes means that two subsequent reaction pairs occur in the same layer.\\
\dag\ The cooling luminosity $L_{\nu}$ scales with temperature $T$, local gravitational acceleration $g$, neutron star radius (in the local rest frame) $R$, and mass fraction $X$ of the respective \ECb\ pair as $L_{\nu} \propto XR^{2}g^{-1}T^{5}$. The temperature scaling assumes $E_x \ll kT$. For further details, see the Supplementary Information S1. The values for $L_{\nu}$ we quote here are for $T = 0.51\,\mathrm{GK}$, $g = 1.85\times 10^{14}\,\mathrm{cm/s^{2}}$,  $R = 12\,\mathrm{km}$, and $X=1$. The existence of the \titanium[56]-\scandium[56] \ECb\ pair depends strongly on nuclear masses. Otherwise, nuclear physics uncertainties of the predicted luminosities are of the order of a factor of 3--4 (see Supplementary Information S5). \\
\ddag\ The transition always occurs at the specified electron chemical potential. The density, which is for a composition consisting of nuclei with a single mass number $A$, will only be approximate for an arbitrary composition.
\end{table}

\renewcommand*{\figurename}{Extended Data Figure}
\renewcommand*{\thefigure}{\arabic{figure}}
\setcounter{figure}{0}

\begin{figure}
\centerline{\includegraphics[width=\doublewide]{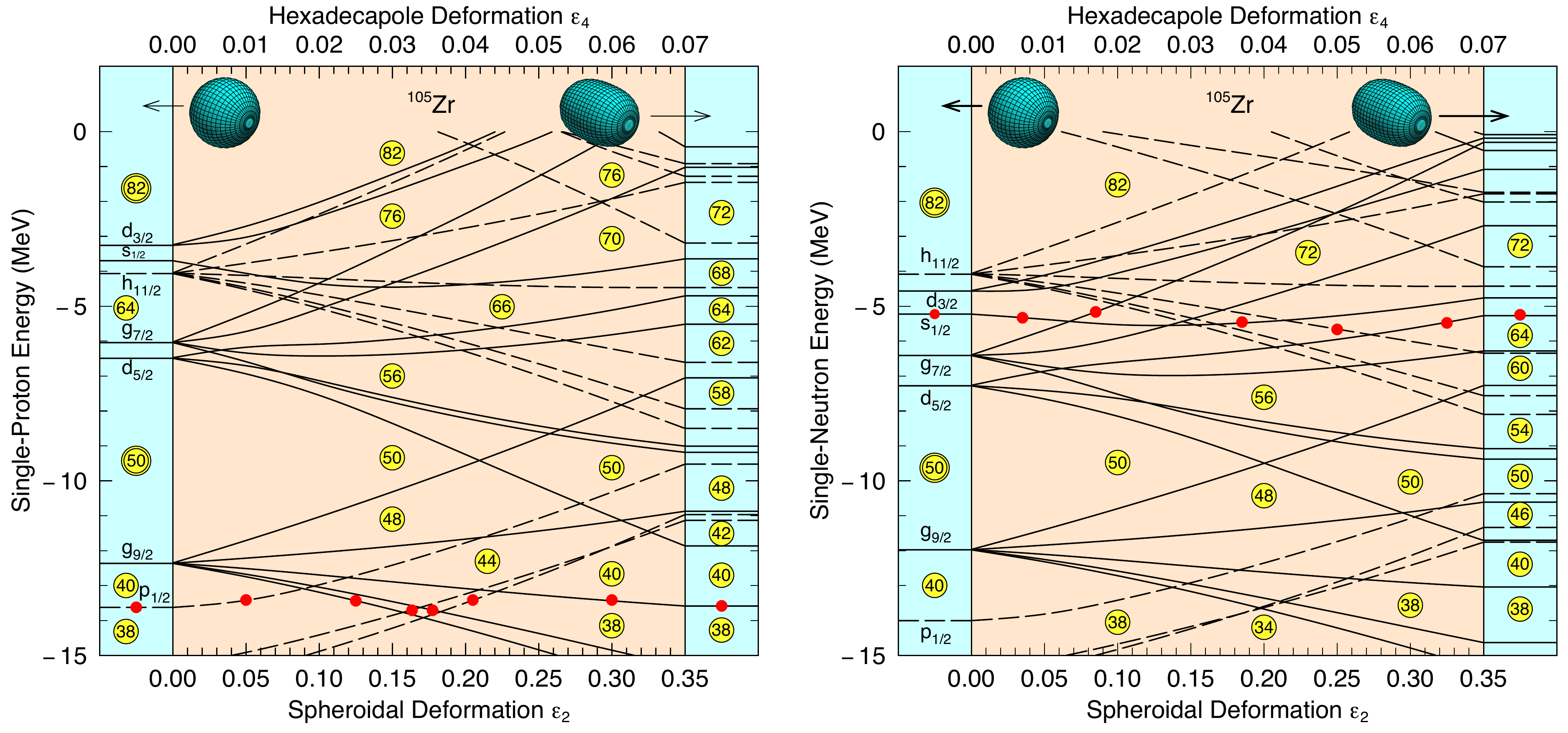}}
\caption{\label{Fig_Nilsson}
\textbf{Calculated proton and neutron single-particle energy levels in $\mathbf{^{105}Zr}$ as functions of nuclear deformation.}
The 40 protons and 65 neutrons in \zirconium[105] fill
all levels up
to the Fermi levels corresponding to these nucleon numbers in the two diagrams (red dots). Levels corresponding to even parity are shown as solid lines,
those corresponding to odd parity as dashed lines. Shell gaps
are characterized by
particularly large separation in energy between two 
adjacent single-particle levels. The numbers of protons or neutrons that occupy
levels up to the shell gap are indicated by circled numbers.
The single-particle levels are shown for a spherical nucleus in
spectroscopic standard notation (left side of each panel), and for a deformation
near the  calculated ground-state shape of \zirconium[105] with quadrupole and hexadecapole shape-parameter values $\epsilon_2=0.333$ and $\epsilon_4=0.06$, respectively\cite{Moller1995}
(right side of each panel). The middle section of each panel shows the change in level energies as
$\epsilon_2$ and $\epsilon_4$ change from  spherical values
$\epsilon_2=\epsilon_4=0$ to deformed values\protect\cite{nilsson55:a}.
The well-known  "magic numbers" 50 and
82 corresponding to particularly large gaps
stand out  at zero deformation\protect\cite{mayer50:a}. 
When the
nuclear shape becomes deformed, the spherical shell gaps disappear
resulting in a large density of levels in the vicinity of the Fermi
level. 
This gives rise to a large
number of states at low excitation in \zirconium[105]. Some of these
states can be populated by strong $\beta^-$ decay transitions from the ground state of 
\yttrium[105]. The situation is similar for  the
electron capture on \zirconium[105] into deformed \yttrium[105]. 
}
\end{figure}

\begin{figure}[htbp]
\centerline{\includegraphics[width=\figwidth]{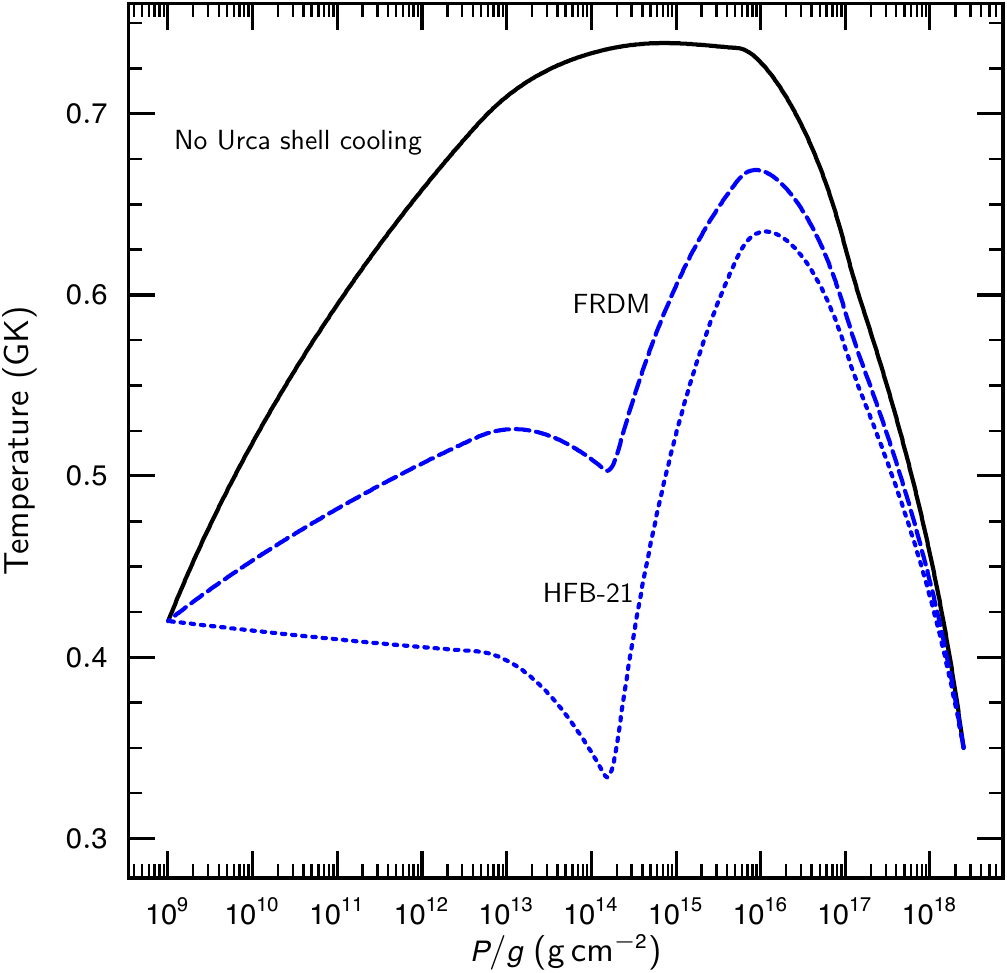}}
  \caption{\textbf{Temperature as a function of depth in the accreted neutron star crust for different Urca shell cooling strengths.} Here we use $P/g \approx \int \rho\,\dif z$ as a proxy for depth.
As a baseline model, we fix the temperature to be $T = 0.42\,\mathrm{GK}$ at $P/g= 10^{9}\,\mathrm{g\,cm^{-2}}$ and $T = 0.35\,\mathrm{GK}$ at the crust-core transition. In the absence of Urca shell cooling, the peak local temperature reaches $0.73\,\mathrm{GK}$ (solid curve)
with the temperature at the superburst ignition depth ($P/g \sim 10^{12}\,\mathrm{g\,cm^{-2}}$) being $0.66\,\mathrm{GK}$. 
With the addition of cooling using the HFB-21 mass model and a superburst ash composition (blue dotted line) a local temperature minimum, $T = 0.33\,\mathrm{GK}$, appears at the location of the Urca shell. Indeed, for these conditions the temperature at the Urca shell is lower than that at the upper boundary, so that a temperature inversion develops.  Even for the much lower Urca shell emissivity of the FRDM mass model (blue dashed line), the temperature at the depth of the superburst ignition is $\lesssim 5\times 10^{8}\,\mathrm{K}$, which is inconsistent with typical superburst ignition conditions\protect\cite{Cumming2006}.  For both mass models, the temperature has a local minimum at the location of the Urca shell.  The steady-state cooling luminosity from the shell is $2\times 10^{35}\, \mathrm{erg\,s^{-1}}$ for the HFB-21 mass model and $1.4\times 10^{35}\,\mathrm{erg\,s^{-1}}$ for the FRDM mass model. As a result, the Urca shell thermally decouples the envelope of light elements from the heating in the deeper crust. 
  }
  \label{f.crust_temps}
\end{figure}

\renewcommand{\theequation}{S\arabic{equation}}
\renewcommand{\thefigure}{S\arabic{figure}}
\renewcommand{\thesection}{S\arabic{section}}
\setcounter{figure}{0}
\setcounter{section}{0}

\section{Identification of \ECb\ pairs and cooling rate calculations}\label{s.pair-identification}
\providecommand*{\dif}{\ensuremath{\mathrm{d}}}
\providecommand*{\enu}{\ensuremath{\varepsilon_{\nu}}}

We identify potential \ECb\ pairs and calculate their neutrino emissivity at a fiducial temperature using a steady-state accreting neutron star crust model\cite{Gupta2007}. The outer crust is thin ($(\Delta R)_{\mathrm{crust}}/R \ll 1$, where $R\approx 12\,\mathrm{km}$ is a representative value for the neutron star radius\cite{Lattimer2007}) so we approximate the crust as a thin plane-parallel layer.  Under this approximation, the pressure of a fluid element increases with time $t$ as $P = \dot{m}gt$, where $\dot{m}$ is the local accretion rate per unit area and $g$ is the local gravitational acceleration.  For the model runs presented here, both $\dot{m}$ and $g$ are set to typical values\cite{strohmayer03:review} for accreting neutron stars, $\dot{m} = 2.6\times10^{4}\,\mathrm{g\,cm^{-2}\,s^{-1}}$ and $g = 1.85\times10^{14}\,\mathrm{cm\,s^{-2}}$. The mass density $\rho$ and electron chemical potential $\mu_{e}$ are determined from an equation of state, described in Ref.~\citen{Brown2009}, that includes contributions from the degenerate, relativistic electrons and the strongly coupled ions. In a significant improvement from Ref.~\citen{Gupta2007}, 
both electron capture transitions and $\beta^-$-decay transitions are included and 
treated on an equal footing. For both types of transitions the $ft$-values are calculated as functions of excitation energy using a quasi-particle random phase approximation\cite{Moller1990}. 
Nuclear masses are taken from experiment\cite{Audi2003} if available, otherwise the predictions of the HFB-21 model\cite{Goriely2010} are used.  

The crust model follows the changes to the composition of an accreted fluid element and the associated energy deposition and neutrino emission through the outer crust of an accreting neutron star using a full nuclear reaction network. The use of a full reaction network is important because the interplay of \ECb\ pairs with subsequent reactions can significantly affect the cooling rates. There are also a few cases where Urca shells for two subsequent \ECb\ pairs overlap. The temperature was kept constant at a fiducial value.
With this approach, cooling \ECb\ pairs can be identified, their approximate depth can be determined, and the cooling rate can be calculated numerically at the chosen temperature
(see Tab.~\ref{Tab_top}). The shell in which the \ECb\ cycles occur has a thickness\cite{Cooper2009}
\begin{equation}\label{e.shell-thickness}
(\Delta R)_{\mathrm{shell}} \approx R\left|\frac{\dif\ln R}{\dif\ln P}\right| \frac{\dif\ln P}{\dif\ln \mu_{e}}\frac{kT}{\mu_{e}} \approx Y_{e}\frac{kT}{m_{\mathrm{u}} g } \sim 1\,\mathrm{m}.
\end{equation}
Here $Y_{e}$ is the electron abundance and $m_{\mathrm{u}}$ is the atomic mass unit.  
Because $(\Delta R)_{\mathrm{shell}}/R \ll 1$, we obtain the 
neutrino luminosity by boosting to the local rest frame and integrating the specific neutrino emissivity $\enu$ over the thin shell:
\begin{equation}\label{e.lum-a}
  L_\nu \approx \int_{\mathrm{shell}} \!\!\enu\, \dif a.
\end{equation}
Here $a$ is the Lagrangian (rest) mass coordinate. We use the relativistic stellar structure equations\cite{Thorne1977} and the one-to-one relation between pressure and $\mu_e$ in the outer crust to transform the integration variable to $\mu_e$,
\begin{equation}
  \label{e.lum-mu}
  L_\nu \approx \frac{4\pi R^2}{g}\left[\int_{\mathrm{shell}} \!\!\enu\frac{\dif P}{\dif\mu_e}\,\dif\mu_e\right].
\end{equation}
Here we have taken the local radial coordinate $R$ and local gravitational acceleration $g = GM/R^2 (1-2GM/Rc^2)^{-1/2}$ as constant over the Urca shell. The term in brackets is determined from our reaction network integration; it is proportional to the mass fraction $X$ of the relevant \ECb\ pair and scales with temperature as $T^5$. This temperature scaling\cite{Tsuruta1970} is valid for $E_x \ll kT$. 
Most \ECb\ pairs listed in Table~\ref{Tab_top} have $E_x = 0$, and of the exceptions, most have $E_x$ of a few tens of keV, which is indistinguishable from $E_x=0$ within model uncertainties. In the absence of more detailed experimental information the assumptions $E_x = 0$ and
$L_\nu \propto T^5$ over all relevant temperatures are therefore reasonable. The only exceptions are the $\beta^{-}$-decays of \calcium[55] and \scandium[57], for which the lowest transitions are predicted to have $E_x=0.27\,\mathrm{MeV}$ and $0.26\,\mathrm{MeV}$, respectively. Both $\beta^{-}$-decays are part, however, of a secondary \ECb\ pair that only contributes weakly to the neutrino emissivity in the respective mass chain. 

To explore the full range of possible compositions\cite{Schatz2001} and to identify all possible \ECb\ pairs that may be activated in neutron star crusts, we investigated 87 different initial compositions, each consisting of a single nuclear species with mass number $A$, where $A$ ranges from 20 to 106. This approach is justified in the outer crust, where electron captures and $\beta^-$ decays are the only reactions and each $A$ chain can be considered independently. In the present work we limit the calculations to the outer crust, and stop the evolution once neutron capture and emission begin to change $A$ of the composition. The types of reactions occurring may then depend on the exact initial composition. We find that reactions changing $A$ typically occur when the neutron mass fraction becomes larger than $10^{-20}$.  A test calculation with X-ray burst ashes shows no evidence for additional strong Urca shells in the inner crust where the neutron mass fraction exceeds $10^{-20}$.

\section{Results for realistic crust compositions}\label{s.realistic-compositions}

For regular X-ray burst ashes\cite{Schatz2001} we find that the \ECb\ pairs \titanium[56]-\scandium[56], \titanium[55]-\scandium[55], and \aluminum[33]-\magnesium[33] dominate and contribute
roughly equally to the cooling rate, with additional contributions from the 
\aluminum[31]-\magnesium[31] and \iron[65]-\manganese[65]
pairs. We assume that the isotopic abundances in the neutron star crust are identical as those produced in the burst and neglect any phase separation\cite{Horowitz2009} and mixing in the liquid layer\cite{Medin2011} that occurs where the crust solidifies.
Obviously, the cooling rate
will be sensitive to the X-ray burst nuclear physics that determines the relative 
proportions of $A=31$, 33, 55, 56, and 65 material in the burst ashes. Superbursts
are predicted to produce ashes with a smaller compositional spread, 
dominated by \iron[56]\cite{Keek2011,Keek2012}.
 In this case cooling is dominated 
by the 
\titanium[56]-\scandium[56]
\ECb\ pair, with a 10\% contribution from a small $A=55$ 
admixture triggering the 
\titanium[55]-\scandium[55]
pair. 

\section{Steady-state thermal transport model}\label{s.transport}

To determine the impact of Urca shell cooling on superburst ignition,
we compute the steady-state crust temperature profile by integrating the relativistic stellar structure equations for temperature and luminosity\cite{Thorne1977} over the crust. The integration does not resolve the individual thin electron capture layers, but instead smooths out the heating and Urca shell cooling over a numerically resolvable thickness.  Full details of the code, along with the equation of state, thermal conductivity, and thermal neutrino emissivities, are described in Ref.~\citen{Brown2009}.
The core and envelope temperatures are taken to match those used in Ref.~\citen{Gupta2007}. The thermal conductivity is computed from the formula appropriate for a crust with an amorphous lattice structure.  This low thermal conductivity causes a steep temperature gradient, and hence high temperatures, in the crust. This model, without any Urca shell cooling, serves as a baseline.  It is similar to the hottest crust models presented in Ref.~\citen{Cumming2006}.  A steadily accreting neutron star will relax to this thermal profile within a few  thermal diffusion timescales, which is of order years\cite{Brown2009}.

We then add a cooling source term to the neutrino emissivity, with a luminosity appropriate for a superburst ash mixture (see Tab.~\ref{Tab_top}), and allow the crust to numerically relax to a steady-state thermal profile.  The shell cooling  results in a dramatically lower temperature in the outer crust (Extended Data Figure~\ref{f.crust_temps}).  For this accretion rate, the neutrino luminosity of the shell is 33\% (23\%) of the total nuclear heating luminosity generated in the deep crust for the HFB-21 (FRDM) mass model. The conductive flux in the layer above the Urca shell, and hence the temperature gradient, is greatly reduced.  For the run using the HFB-21 mass model, the luminosity in this layer is negative, so that a temperature inversion develops. The Urca shell effectively reduces the thermal flux reaching the \carbon-rich layer from the deep crust.  As a result, the high temperatures necessary to ignite \carbon\ cannot be the result of deep crustal heating, even under the best of circumstances.

\section{Influence of nuclear deformation}\label{s.deformation}
The Urca shell cooling process depends strongly on the transition
strengths of the electron capture and $\beta^{-}$-decay processes within an \ECb\
pair to low lying states, sometimes including the ground state. 
 It is therefore closely linked to the level structure of neutron-rich
nuclei. This structure depends on nuclear shape.  It is
unusual for spherical neutron-rich nuclei to exhibit strong electron
capture or $\beta^{-}$-decay to very low-lying daughter states. On the other
hand, deformed nuclei have a significantly altered single-particle
structure as the non-spherical shape splits the highly degenerate spherical 
levels into a large number of doubly degenerate levels spaced more 
uniformly and densely in the vicinity of the Fermi surface (Extended Data Figure~\ref{Fig_Nilsson}). 
This results in a large number of low-lying
energy levels
in deformed nuclei, and in particular in odd-A nuclei.
Therefore  strong, allowed transitions to
low-energy daughter states are much more likely.
This effect of deformation in our QRPA model has been demonstrated previously and agrees
well with experimental data
{\cite{Krumlinde1984}\textsuperscript{,}\cite{Moller1990}}.  The effect is mainly responsible for 
the high
cooling rates found in this work and is the reason why most of the strong
\ECb\ pairs are found in regions of the nuclear chart where large
deformations are predicted.

\section{Uncertainties in the nuclear physics}\label{s.uncertainties}
Our phase space approximation used in the calculation of electron capture rates and $\beta^-$ decay rates  
provides an accuracy for the resulting neutrino luminosity of 20\%. In addition there are significant nuclear physics uncertainties. The Quasi Particle Random Phase Approximation method used to calculate the strengths of individual electron capture and $\beta^-$ transitions has significant uncertainties when predicting the strength of individual transitions at low excitation energy\cite{Cole2012}. However, it is the only available global approach that can be applied to all nuclei of relevance. For 5 \ECb\ pairs in Table~\ref{Tab_top}, \magnesium[29]-\sodium[29], \aluminum[31]-\magnesium[31], \aluminum[33]-\magnesium[33], \chromium[57]-\vanadium[57], and \vanadium[57]- \titanium[57], there are experimental data available on the $ft$-value for the ground-state-to-ground-state $\beta^-$ decay. Comparison with the theoretical prediction indicates an average deviation of about a factor of 3. An exception is the \aluminum[31]-\magnesium[31] pair, where the experimental  limit of $\log ft > 6.9$ for the ground state 
$\beta^-$ decay of \magnesium[31] differs by more than two orders of magnitude from the theoretical prediction of $\log ft = 4.4$. This is not unexpected since the experimental data indicate\cite{Marechal2005} that this \ECb\ pair lies on different sides of the border around the so-called island of inversion, where ground state configurations of nuclei change drastically with small changes in proton or neutron number because of large changes in deformation. With \aluminum[31] lying outside of the island of inversion, and 
\magnesium[31] lying inside, the states near the Fermi surface, if expanded in a spherical basis, are dominated by different major shells, which reduces the transition strength dramatically. Our model assumes similar shapes for parent and daughter nuclei, which is a good approximation for most nuclei, though not in this exceptional case.  Nevertheless, the \aluminum[31]-\magnesium[31] pair does not play a critical role in our study, and the uncertainties of a factor of 3 found for the other \ECb\ pairs can be regarded as a more typical indicator of the theoretical model uncertainty. Further measurements of ground state $ft$-values for neutron rich nuclei, especially for the \ECb\ pairs listed in Table~\ref{Tab_top}, would further reduce nuclear physics uncertainties. 

Nuclear 
masses are also not known experimentally for many of the \ECb\ pairs and need to be 
predicted with nuclear mass models. To 
test the dependence on nuclear masses, we performed  calculations using the FRDM
mass model\cite{Moller1995} instead of the HFB-21 mass model\cite{Goriely2010}. Overall the results are comparable and cooling rates for most \ECb\ pairs agree within 20\%. One important exception is the 
\titanium[56]-\scandium[56] pair, which does not form an effective \ECb\ pair when the FRDM masses are used.
The reason is that the $|\Qec|$ values predicted by FRDM have a much larger staggering between odd and even charge numbers, for a given $A$, than those predicted by the HFB-21 mass model ($4.1\,\mathrm{MeV}$ for FRDM versus $1.0\,\mathrm{MeV}$ for HFB-21).  Since the predicted 
excitation energy of the transition from \scandium[56] to \calcium[56] is only $3.4\,\mathrm{MeV}$, a model using the FRDM masses finds that electron capture on \scandium[56]  occurs as soon as electron capture on \titanium[56]
becomes energetically possible.  This strongly limits cycling in the \titanium[56]-\scandium[56] \ECb\ pair and reduces the neutrino luminosity from a crust composed of superburst ashes by roughly 90\% compared to a calculation using the HFB-21 mass model.
Measurements at radioactive beam facilities of the  
masses of \calcium[56] and \scandium[56] and the excitation energy of the allowed transitions are necessary to identify the strength of the cooling in superburst ashes. Other larger 
changes when calculating cooling rates with the FRDM mass model are the \ECb\ pairs \chromium[63]-\vanadium[63] and \zirconium[105]-\yttrium[105] that are reduced by more than a factor of 2. On the other hand, the \magnesium[31]-\sodium[31] pair becomes a very strong cooling pair, on par with \magnesium[29]-\sodium[29]. 

Despite the considerable theoretical uncertainties in predicting the strength of individual \ECb\ pairs, our main conclusions are robust owing to the steep temperature dependence of the cooling rate and the fact that a realistic composition will be spread over a range of mass numbers and will therefore activate a range of \ECb\ pairs. 


\renewcommand{\refname}{Supplemental References}

\end{document}